\documentclass[runningheads]{llncs}

\usepackage[T1]{fontenc}
\usepackage{graphicx}
\usepackage{comment}
\usepackage{amsmath,amssymb}
\usepackage{color}
\usepackage{url}
\usepackage{hyperref}
\usepackage{multirow}
\usepackage{array}

\newif\ifreview
\reviewfalse

\ifreview
	\usepackage{lineno}

	\linenumbers
\fi

\begin{document}

%%%%%%%%%%%%%%%%%%%%% Add submission id, track, and title. %%%%%%%%%%%%%%%%%%%%%

% TODO: Please insert your submission number here
\def\SubNumber{000}

% TODO: Please uncomment the track this paper will be submitted to, comment all other lines
\def\GCPRTrack{Main Track}
%\def\GCPRTrack{Special Track: Pattern recognition in the life and natural sciences}
%\def\GCPRTrack{Special Track: Photogrammetry and remote sensing}
%\def\GCPRTrack{Young Researcher's Forum}
%\def\GCPRTrack{Fast Review Track}

% TODO: Replace with your title
\title{HistDiST: Histopathological Diffusion-based Stain Transfer}
% You can use \thanks for acknowledgment as in: 
%\title{Title\thanks{XXX}}
%Do not add any acknowledgment to the draft 
% version that is used for the review process.  

\ifreview
	% ANONYMOUS SUBMISSION FOR REVIEW
	% DO NOT MODIFY these for the draft version that is used for the review process.
	\titlerunning{GCPR 2025 Submission \SubNumber{}. CONFIDENTIAL REVIEW COPY.}
	\authorrunning{GCPR 2025 Submission \SubNumber{}. CONFIDENTIAL REVIEW COPY.}
	\author{GCPR 2025 - \GCPRTrack{}}
	\institute{Paper ID \SubNumber}
\else

	\author{Erik Großkopf$^*$,
Valay Bundele$^*$,
Mehran Hosseinzadeh$^*$,
Hendrik P.A. Lensch
}
	
	\authorrunning{Erik Großkopf,
Valay Bundele,
Mehran Hosseinzadeh, Hendrik P.A. Lensch}
	
	\institute{University of Tübingen, Germany\\
\email{erik.grosskopf@student.uni-tuebingen.de, \{valay.bundele, mehran.hosseinzadeh, hendrik.lensch\}@uni-tuebingen.de}}
\fi

\maketitle              % typeset the header of the contribution

\def\thefootnote{*}\footnotetext{These authors contributed equally to the work.}

\begin{abstract}
Hematoxylin and Eosin (H\&E) staining is the cornerstone of histopathology but lacks molecular specificity. While Immunohistochemistry (IHC) provides molecular insights, it is costly and complex, motivating H\&E-to-IHC translation as a cost-effective alternative. Existing translation methods are mainly GAN-based, often struggling with training instability and limited structural fidelity, while diffusion-based approaches remain underexplored. We propose HistDiST, a Latent Diffusion Model (LDM) based framework for high-fidelity H\&E-to-IHC translation.
HistDiST introduces a dual-conditioning strategy, utilizing Phikon-extracted morphological embeddings alongside VAE-encoded H\&E representations to ensure pathology-relevant context and structural consistency. To overcome brightness biases, we incorporate a rescaled noise schedule along with $v$-prediction, enforcing a zero-SNR condition at the final timestep. During inference, DDIM inversion preserves the morphological structure, while an $\eta$-cosine noise schedule introduces controlled stochasticity, balancing structural consistency and molecular fidelity. Moreover, we propose Molecular Retrieval Accuracy (MRA), a novel pathology-aware metric leveraging GigaPath embeddings to assess molecular relevance. Extensive evaluations on MIST and BCI datasets demonstrate that HistDiST significantly outperforms existing methods, achieving a 28\% improvement in MRA on the H\&E-to-Ki67 translation task, highlighting its effectiveness in capturing true IHC semantics. The code is available at https://github.com/ErikGro/HistDiST.

\end{abstract}

\section{Introduction}
Histopathology is essential for disease diagnosis, revealing tissue structure and cellular morphology. While Hematoxylin and Eosin (H\&E) staining is cost-effective and highlights structural details, it lacks molecular specificity. In contrast, Immunohistochemistry (IHC) provides molecular insights but is expensive and time-consuming. Automated H\&E-to-IHC translation offers a scalable solution by inferring molecular details from tissue morphology, reducing costs, and enhancing diagnostic consistency. Deep learning methods, particularly GANs, 
GANs have shown promise in this task. Li et al. ~\cite{li2023adaptive} use adaptive supervised PatchNCE loss to preserve structure, while  MDCL ~\cite{wang2024mix} employs a conditional GAN with multi-domain contrastive learning to enhance cross-domain alignment. Pathology-specific approaches include patch-level feature extraction with multiple instance learning ~\cite{li2024virtual} and HER2 scoring enhancement via nuclei density estimation ~\cite{peng2024advancing}. However, GANs still face mode collapse and require complex loss designs to maintain consistency in pathology screenings.
% Histopathology is critical for disease diagnosis, offering insights into tissue architecture and cellular morphology. Hematoxylin and eosin (H\&E) staining is popular due to its affordability and ability to highlight structural details but lacks molecular specificity. Immunohistochemistry (IHC) provides essential molecular information but is expensive, time-consuming, and prone to inconsistency. 
% Automated translation of H\&E-stained images into IHC representations offers a promising solution by inferring molecular details directly from tissue morphology. Such methods can streamline pathology workflows, reduce costs, and improve diagnostic consistency. Deep learning methods, particularly Generative Adversarial Networks (GANs), have shown potential in this task. \hl{a number cannot be the subject of a sentence. Use e.g. citet} For example, Li et al. ~\cite{li2023adaptive} employ an adaptive supervised PatchNCE loss to maintain structural integrity, while ~\cite{chen2024pathological} improves translation fidelity through a multi-branch discriminator with prototype consistency. Pathology-specific approaches have also emerged: ~\cite{li2024virtual} incorporates a patch-level pathological feature extractor with multiple instance learning for virtual staining, and ~\cite{peng2024advancing} enhances HER2 scoring using a nuclei density estimator and an auxiliary alignment branch. Despite these advances, GAN-based methods often suffer from mode collapse and require meticulous loss design to preserve pathological consistency.

Recently, diffusion models (DMs) ~\cite{ho2020DDPM} have emerged as a powerful alternative, offering superior stability and synthesis quality. Latent Diffusion Models (LDMs) ~\cite{rombach2022LDM} reduce the computational burden of DMs by operating in the latent space of pretrained autoencoders. LDM-based image editing methods, such as InstructPix2Pix ~\cite{brooks2023instructpix2pix} for instruction-based editing and ControlNet ~\cite{zhang2023adding} for conditional generation, offer structured control. Inference-time optimization techniques using DDIM Inversion~\cite{tumanyan2023DDIM} and Delta Denoising Score~\cite{hertz2023DDS} further enhance structural fidelity.
%Additionally, LDM-based image editing methods have shown promise. InstructPix2Pix ~\cite{brooks2023instructpix2pix} fine-tunes Stable Diffusion for instruction-based image editing using large-scale image-text pairs, while ControlNet introduces conditional inputs (e.g., sketches, edge maps) to precisely constrain generation. Inference-time optimization techniques, such as the use of inversion noise  ~\cite{tumanyan2023DDIM} and Delta Denoising Score~\cite{hertz2023DDS}, further enhance structural fidelity and control in a training-free setup. 
Although there has been quite some work on applying LDMs for natural image editing, the application of LDMs for H\&E-to-IHC stain translation remains largely unexplored. Existing diffusion-based methods \cite{dubey2024vims,yan2025versatile} mainly use text prompts or unpaired signal vector conditioning, but often fail to ensure structural and semantic consistency, critical for accurate pathology translation.

To address these limitations, we propose HistDiST, an LDM-based framework for high-fidelity H\&E-to-IHC stain translation.  Although LDM offers strong generative capabilities, its direct application to stain translation faces three key challenges: (i) lack of explicit morphological conditioning, (ii) brightness bias from conventional diffusion noise schedules, and (iii) insufficient structural preservation at inference. These limitations hinder the model’s ability to capture complex morphology–molecular relationships and lead to inaccurate brightness distributions that obscure diagnostically relevant molecular features.

HistDiST overcomes these challenges through three key contributions. First, we employ a dual-conditioning strategy that integrates pathology-specific priors into denoising process. Specifically, Phikon-based embeddings~\cite{filiot2023scaling} are injected via cross-attention within the U-Net, while VAE-encoded H\&E representations are concatenated with latent noise input. 
%First, a dual-conditioning strategy integrates Phikon-based ~\cite{filiot2023scaling} pathology-specific embeddings into the U-Net’s cross-attention layers and concatenates VAE-encoded H\&E representations with the input latent noise. 
This ensures that pathology-relevant context guides denoising for precise molecular reconstruction.
%This ensures that morphological context informs the denoising process from the earliest stages, enabling precise molecular reconstruction. 
Second, to address brightness inconsistencies, we incorporate a rescaled noise schedule and trailing timesteps ~\cite{lin2024common} along with $v$-prediction ~\cite{karras2022elucidating}, enforcing zero terminal SNR. 
Third, we perform joint training for unconditional H\&E generation and H\&E-to-IHC translation, a critical step that enables the use of DDIM inversion ~\cite{song2020denoising} for structure preservation at inference. 
%DDIM inversion preserves structural consistency by mapping the input H\&E image to its corresponding latent noise vector, allowing the model to reconstruct IHC images while retaining key morphological features.
However, while inversion effectively anchors structure, the subsequent DDIM denoising—being fully deterministic—restricts generative flexibility. As a result, the model fails to capture subtle but diagnostically relevant molecular variations in the translated IHC image.

To overcome this limitation, we introduce an $\eta$-cosine noise schedule that progressively increases stochasticity during DDIM denoising. This modification enables controlled exploration of molecular solution space while preserving morphological consistency. The resulting outputs show improvements in synthesis diversity and fidelity, as reflected by lower FID scores.
%Although DDIM inversion helps with structure preservation, the deterministic nature of subsequent DDIM sampling restricts generative flexibility, limiting the model's ability to capture subtle molecular variations critical for accurate IHC synthesis. To mitigate this, we incorporate an $\eta$-cosine noise schedule, progressively increasing the noise scale during denoising to enable stochastic exploration. This significantly improves FID scores while preserving both morphological structures and molecular fidelity in the translated IHC images. 
Moreover, we propose Molecular Retrieval Accuracy (MRA), a novel feature-based metric that quantifies molecular fidelity using cosine similarity. Experiments on MIST and BCI datasets demonstrate that HistDiST outperforms state-of-the-art methods across both qualitative and quantitative benchmarks, achieving superior visual realism, structural preservation, and molecular accuracy. Notably, HistDiST yields 28\% improvement in MRA on H\&E-to-Ki67 translation, demonstrating highly accurate IHC synthesis.
\section{Related Work}

\subsection{Image Generation using Diffusion Models}

Diffusion models (DMs) have emerged as a powerful alternative to GANs for image generation, offering greater diversity and training stability by learning the data distribution through iterative denoising processes~\cite{ho2020DDPM}. These models operate either directly in the data space or in a compressed latent space, as in Latent Diffusion Models (LDMs)\cite{rombach2022LDM}, enabling more efficient computation. Building on their success in generative tasks, conditional diffusion models have been developed to allow guided image synthesis for applications such as image-to-image translation \cite{nichol2021glide} and text-based image editing ~\cite{brooks2023instructpix2pix}. Conditioning is typically achieved by injecting external information into the denoising network. For instance, Stable Diffusion~\cite{rombach2022LDM} and InstructPix2Pix~\cite{brooks2023instructpix2pix} leverage text prompts mapped to a shared embedding space with images, while ControlNet~\cite{zhang2023controlnet} enhances this framework by incorporating spatial guidance, enabling fine-grained structural control. In addition, inference-time optimization techniques—such as DDIM inversion~\cite{tumanyan2023DDIM} and Delta Denoising Score (DDS)~\cite{hertz2023DDS}—have been introduced to improve structural fidelity and control in a training-free manner.

\subsection{Stain Transfer in Computational Pathology}
The generation of virtually stained images given a source staining type image is an active field of research, aiming to alleviate the lack of useful IHC counterparts to H\&E images and meanwhile preventing the challenges caused by re-staining or re-sampling tissues. Generative Adversarial Networks (GANs) have been widely employed for H\&E-to-IHC translation, with many approaches focusing on preserving structural and pathological integrity. For example, Li et al.\cite{li2023adaptive} introduce an adaptive supervised PatchNCE loss to maintain structure, while Chen et al.\cite{chen2024pathological} improve translation fidelity via a multi-branch discriminator with prototype consistency. Other methods incorporate domain-specific enhancements: Li et al.\cite{li2024virtual} propose a patch-level pathological feature extractor with multiple-instance learning, and Peng et al.\cite{peng2024advancing} refine HER2 scoring using a nuclei density estimator and auxiliary alignment branch. 
%Domain knowledge and contrastive learning also play key roles in improving stain transfer. Wang et al.\cite{wang2024mix} apply contrastive learning to align intra- and inter-domain features via anchor-matching of mixed-domain patches, while Ma et al.\cite{ma2024dsff} develop DSFF-GAN, introducing a dual structure–color similarity loss and a specialized DSFF block for more reliable translation. 
Wang et al.\cite{wang2024mix} incorporate Multi-Domain Contrastive Learning (MDCL) within a conditional GAN to enhance cross-domain alignment, and Qu et al.\cite{qu2024advancing} utilize a multi-magnification processing strategy with an attention module to extract fine-grained features while minimizing information loss. Despite their effectiveness, GAN-based methods often struggle with mode collapse and require carefully crafted loss functions to ensure pathological accuracy.

Recently, diffusion models have gained attention for their stability and high-quality synthesis in natural image generation. Their application in virtual staining, however, remains limited. He et al.\cite{he2024pst} leverage diffusion models augmented with Schrödinger bridges and structural constraints to achieve stain transfer, though standard evaluation metrics are not reported. Shen et al.\cite{shen2023staindiff} employ diffusion solely for style transfer, without addressing cross-stain learning. Text-conditioned diffusion has also been explored: Dubey et al.\cite{dubey2024vims} utilize a latent diffusion model guided by text prompts, but their reliance on pixel-aligned datasets limits practical applicability. Yan et al.\cite{yan2025versatile} propose a unified dual-encoder diffusion framework for multi-stain translation across diverse stain types (H\&E, MT, PAS, PASM). Nonetheless, the use of diffusion models for H\&E-to-IHC stain translation remains underexplored, with a lack of methods that explicitly target pathological consistency and structural fidelity in this challenging setting.
%Some works integrate contrastive learning and advanced feature extraction techniques within diffusion models. For example, ~\cite{zhang2024high} introduces a bidirectional contrastive learning loss to enhance patch-wise consistency while mitigating misalignment. 

% \input{sections/3_method}
\section{Methodology}

\subsection{Preliminaries}
Diffusion models generate data by learning to reverse a stochastic forward process that gradually corrupts an input sample \( x_0 \) with Gaussian noise over \( T \) steps, following a variance schedule \( \beta_t \): 

\begin{equation}
q(x_t | x_{t-1}) = \mathcal{N}(x_t; \sqrt{1 - \beta_t} x_{t-1}, \beta_t I).
% \]
\end{equation}

For large \( T \), this process converges to an isotropic Gaussian distribution. The model approximates the reverse process:
% \[
\begin{equation}
p_\theta(x_{t-1} | x_t) = \mathcal{N}(x_{t-1}; \mu_\theta(x_t, t), \Sigma_\theta(x_t, t)),
% \]
\end{equation}

where \( \mu_\theta \) and \( \Sigma_\theta \) are learnable functions. Training involves predicting the noise component \( \epsilon_\theta(x_t, t) \) by minimizing:

% \[
\begin{equation}
L = \mathbb{E}_{x_0, t, \epsilon} \left[ \| \epsilon - \epsilon_\theta(x_t, t) \|^2 \right],
% \]
\end{equation}

where \( \epsilon \sim \mathcal{N}(0, I) \) and $\epsilon_\theta$ is a neural network parameterized by $\theta$. Latent Diffusion Models (LDMs) enhance computational efficiency by operating in a compressed latent space, where an encoder \( E \) maps images \( x_0 \) to latent representations \( z_0 = E(x_0) \).

\subsection{HistDiST}

\begin{figure}[t]
    \centering
    \includegraphics[width=0.45\textwidth, angle=90]{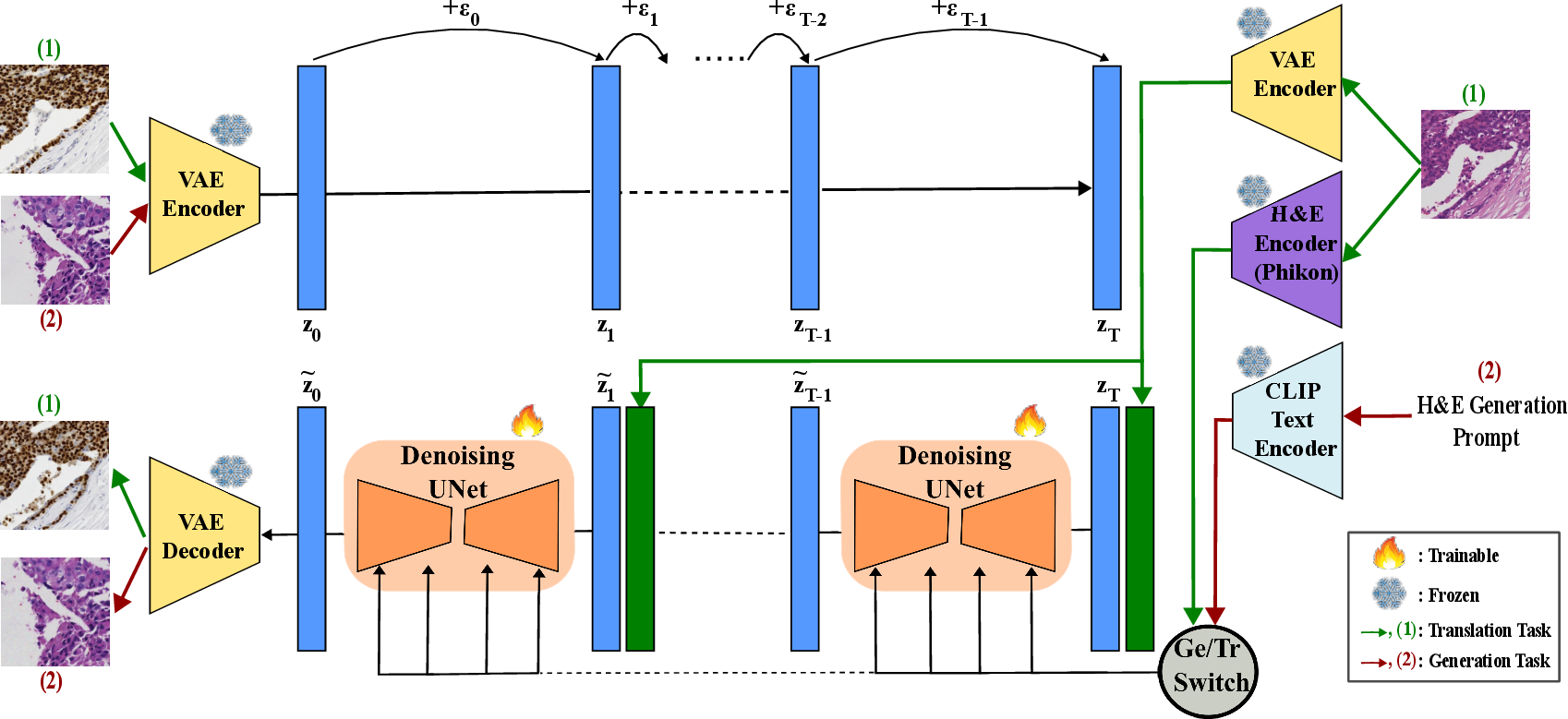}
    \caption{HistDiST training pipeline, showing H\&E generation (red arrows, label (2)) conditioned on CLIP text embeddings, and H\&E-to-IHC translation (green arrows, label (1)) guided by Phikon embeddings and VAE-encoded H\&E features. The VAE encoder maps images to latent space, where noise is added and later denoised by the U-Net. The Ge/Tr switch selects between generation and translation tasks, with each (numbered input, color-coded pathway) independently followed.    
    % HistDiST training pipeline, showcasing H\&E generation, in red arrows and using numbered label (1), conditioned on CLIP text embedding and H\&E-to-IHC translation, in green arrows and using numbered label (2), guided by Phikon-extracted embeddings and VAE-encoded H\&E features. The VAE encoder maps images to latent space, where noise is added and later denoised by the U-Net. The Ge/Tr switch selects between generation and translation, with each numbered input and color-coded pathway independently followed for its respective task.
    }
    \label{fig:yourimage}
\end{figure}
% We present \textbf{HistDiST}, a novel framework that harnesses pre-trained Stable Diffusion for high-fidelity, image-conditioned H\&E-to-IHC stain translation. Stable Diffusion, a text-conditioned latent diffusion model (LDM), employs a U-Net denoising network with residual, self-attention, and cross-attention blocks, supporting multi-modal conditioning with inputs such as text, images, and structural priors. To enable DDIM inversion for structural preservation during inference, we jointly train the model for both unconditional H\&E generation and H\&E-to-IHC translation.
% We present HistDiST, a novel framework that leverages pre-trained Stable Diffusion for high-fidelity H\&E-to-IHC translation. Stable Diffusion, a text-conditioned latent diffusion model (LDM), uses a U-Net denoising network with residual connections and attention mechanisms for multi-modal conditioning. HistDiST is jointly trained for unconditional H\&E generation and H\&E-to-IHC translation, enabling DDIM inversion for structural preservation during inference. The VAE remains frozen, while the modified U-Net is fine-tuned on paired H\&E-IHC data.
We present HistDiST, a novel framework leveraging LDMs for high-fidelity H\&E-to-IHC translation. LDMs operate in a variational autoencoder (VAE) latent space and use a U-Net denoising network with residual connections, self-attention, and cross-attention mechanisms. HistDiST achieves superior structural and molecular accuracy through: (1) dual-conditioning (2) v-prediction with rescaled noise schedules, and (3) DDIM inversion followed by DDIM denoising with $\eta$-cosine scheduling, balancing determinism with controlled stochasticity.

\paragraph{Dual Conditioning:} To capture the intricate relationship between tissue morphology and molecular expression, HistDiST employs a \textit{dual-conditioning} strategy. First, Phikon ~\cite{filiot2023scaling}, a self-supervised transformer trained on 40 million histology images, extracts morphological feature embeddings from the input H\&E-stained image. These embeddings are injected into the cross-attention layers of the U-Net, enabling the model to integrate pathology-relevant context throughout the denoising process. Second, inspired by InstructPix2Pix ~\cite{brooks2023instructpix2pix}, we modify the U-Net by applying additional input channels that allow VAE-encoded H\&E latents to be concatenated with the latent noise vector at the U-Net input. This modification provides structural guidance from the earliest denoising stages, enabling IHC generation with high structural fidelity and molecular accuracy.

\paragraph{Joint Training:} Fig. \ref{fig:yourimage} shows the training pipeline of HistDiST. To enable DDIM inversion for structure preservation at inference, HistDiST is trained jointly on unconditional H\&E generation and H\&E-to-IHC translation.
%We train HistDiST jointly on unconditional H\&E generation and H\&E-to-IHC translation to enable DDIM inversion at inference, allowing the model to map H\&E images to their latent noise and preserve morphological structure in the generated IHC output. 
%H\&E generation is conditioned on CLIP text embeddings, while IHC translation uses the same dual-conditioning strategy. 
While the VAE remains frozen, the U-Net is fine-tuned on paired H\&E-IHC data.
A key limitation in standard diffusion models is that common noise schedules fail to enforce zero SNR at the final timestep, causing residual low-frequency information to be leaked during training ~\cite{lin2024common}. However, at inference, the model starts from pure Gaussian noise, leading to a mismatch between training and inference, resulting in constrained brightness levels. HistDiST addresses this by rescaling the noise schedule under a variance-preserving formulation ~\cite{lin2024common} to enforce zero terminal SNR and enable more accurate brightness distribution in outputs.
%due to common noise schedules, which fail to enforce a zero signal-to-noise ratio (SNR) at the final timestep. This discrepancy leads to incongruent behavior between training and inference phases, where the model learns to rely on residual low-frequency information during training, resulting in generated images constrained to medium brightness levels during inference. To resolve this, HistDiST incorporates key refinements, drawing on recent insights from ~\cite{lin2024common}. Specifically, we rescale the noise schedule under a variance-preserving formulation to enforce zero terminal SNR, aligning training and inference behavior and enabling the generation of images across a diverse brightness range.

Moreover, when SNR is zero, traditional epsilon ($\epsilon$) prediction becomes trivial and does not provide meaningful learning signals. To overcome this, HistDiST adopts a $v$-prediction and $v$-loss framework as proposed in ~\cite{karras2022elucidating}. The velocity vector is defined as:
\begin{equation}
% \[
v_t = \sqrt{\bar{\alpha}_t} \, \epsilon - \sqrt{1 - \bar{\alpha}_t} \, x_0,
% \]
\end{equation}
where $x_0$ is the clean latent representation (noise-free image), $\epsilon$ is the Gaussian noise added during the forward diffusion process, and $\bar{\alpha}_t$ represents the cumulative product of the noise schedule parameters up to timestep $t$. The corresponding \textit{v-prediction} loss function is:
\begin{equation}
    \mathcal{L} = \lambda_t \| v_t - \tilde{v}_t \|_2^2,
\end{equation}

where $v_t$ is the true velocity, $\tilde{v}_t$ is the velocity predicted by the network, and $\lambda_t$ is a timestep-dependent weighting factor. Finally, HistDiST employs trailing timestep selection, prioritizing later timesteps during training to align with inference, where model starts from pure Gaussian noise. This enables generation of images with diverse brightness levels and helps preserve molecular features.

\begin{figure}[t]
    \centering
    \begin{tabular}{m{0.5\textwidth}@{}m{0.4\textwidth}}
        \includegraphics[width=0.27\textwidth, angle=90]{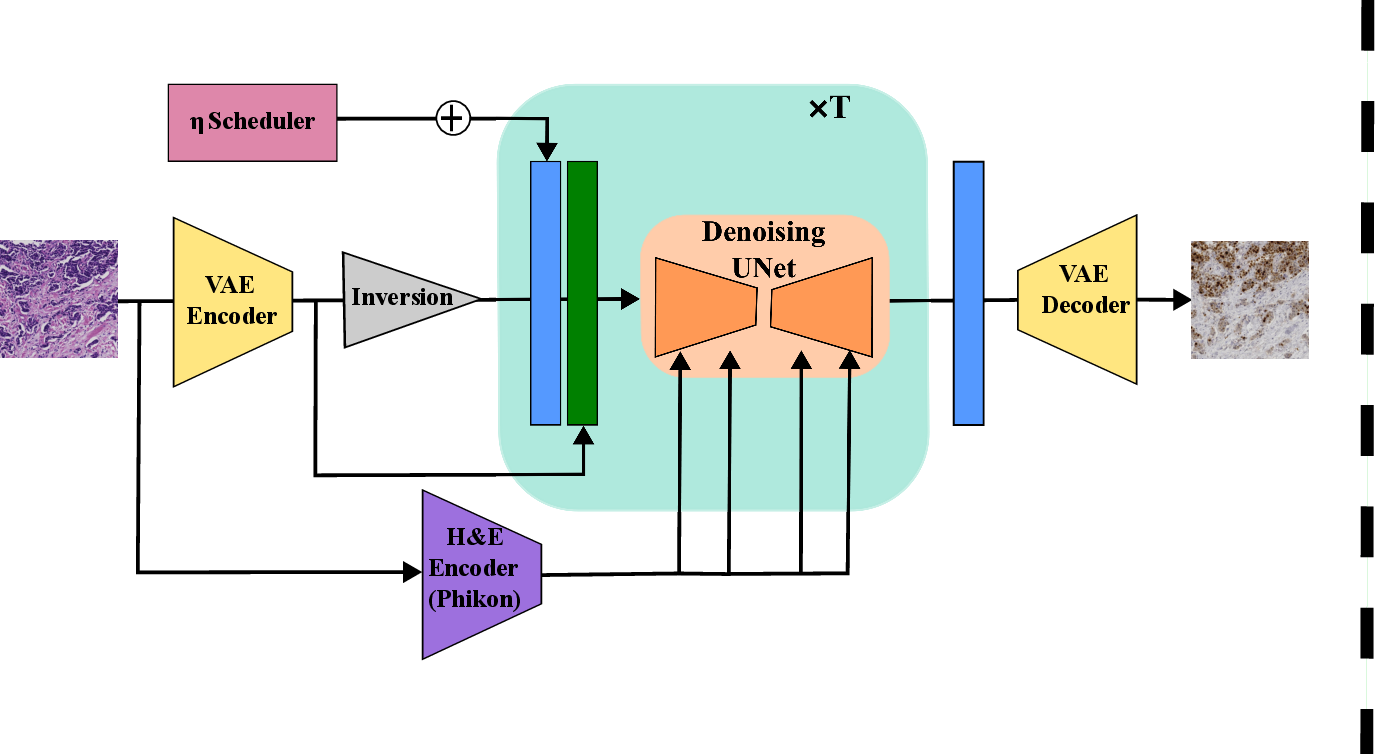} &
        \includegraphics[width=0.4\textwidth]{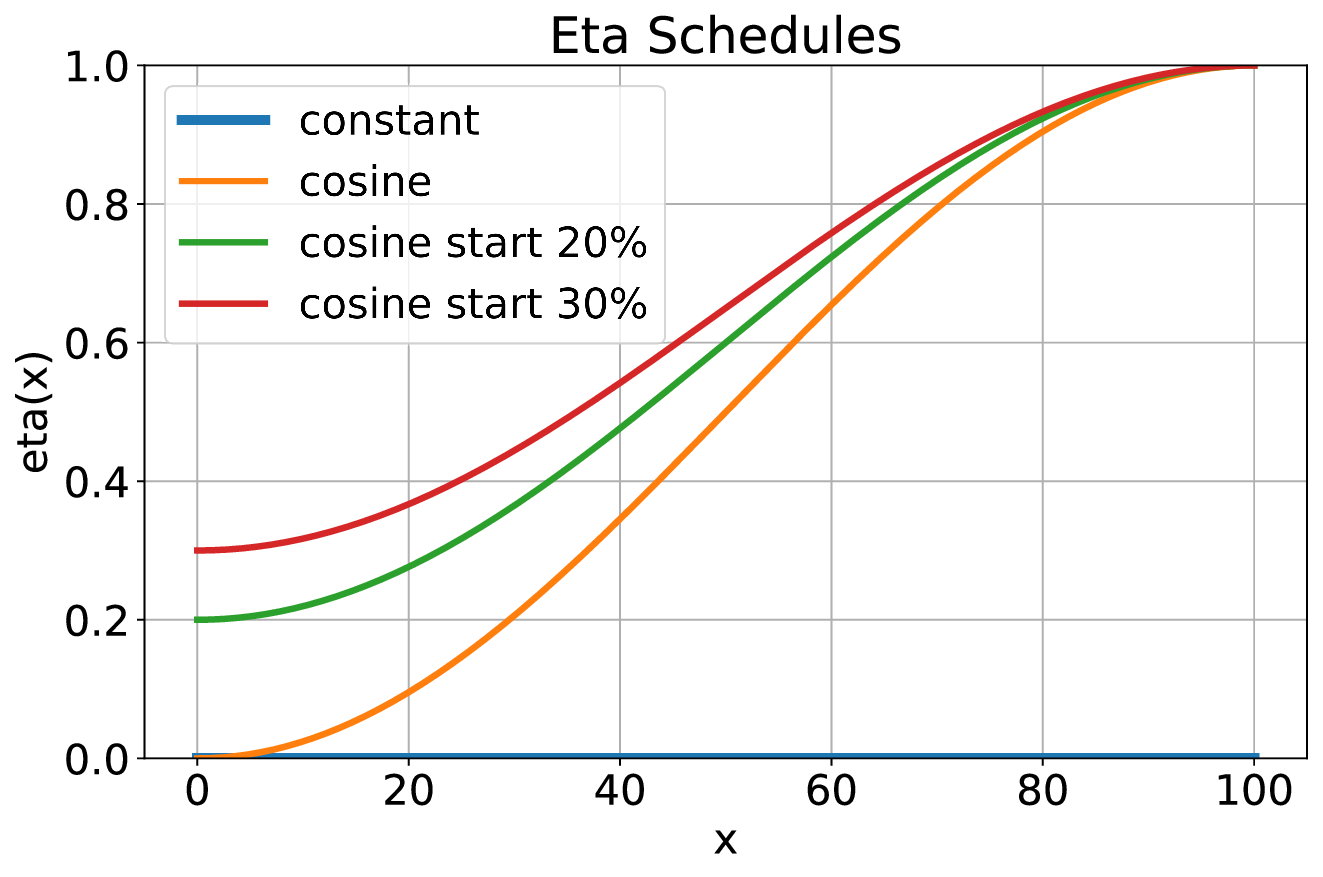} \\
    \end{tabular}
    \caption{HistDiST inference pipeline (Left): VAE encoder maps H\&E image to latent space, where DDIM inversion derives noise latent and $\eta$-noise scheduling injects noise at different timesteps during denoising. The U-Net, conditioned on Phikon embeddings, refines the features, and the VAE decoder generates the final IHC output. (Right) $\eta$-schedules, with cosine-start schedules optimizing FID and structural preservation.}
    \label{fig:inference}
\end{figure}

\paragraph{Inference:} Fig. \ref{fig:inference} (Left) shows the inference pipeline. Preserving the \textit{morphological structure} of input images is essential for H\&E-to-IHC stain translation. To achieve this, we employ DDIM inversion ~\cite{song2020denoising} to map input H\&E representation $z_0$ to its corresponding noise vector $z_T$. This inversion process effectively "reverses" the generative path, allowing us to find the specific noise that encodes structural elements of the input. By using the inverted noise $z_T$ and then sampling with the trained HistDiST, we obtain translated images that maintain the original morphological features while rendering them in the target IHC modality.

The inversion is performed via deterministic updates:
\begin{equation}
z_{t+1} = \sqrt{\alpha_{t+1}} \left( \frac{z_t - \sqrt{1 - \alpha_t} \, \epsilon_\theta(z_t, t, c_T)}{\sqrt{\alpha_t}} \right) 
+ \sqrt{1 - \alpha_{t+1}} \, \epsilon_\theta(z_t, t, c_T),
\end{equation}
where $\alpha_t$ is the cumulative noise schedule, $\epsilon_\theta$ is the predicted noise, and $c_T$ is the text-based conditioning embedding. 
%\hl{lots of italic in the upcoming sentences? Is it necessary?} 
While using this inverted noise as the starting point for translation successfully preserved structural features, we observed a increase in FID scores (Table \ref{tab:image_translation_ablation} (right), $\eta$=0), indicating that deterministic DDIM denoising restricted generative flexibility and hindered the model's ability to capture molecular diversity essential for accurate IHC synthesis.

To address this, we incorporated an $\eta$-cosine noise schedule that introduces \textit{controlled stochasticity} during denoising. Unlike standard deterministic DDIM sampling, our schedule progressively increases the noise scale $\eta_t$ from a non-zero initial value following a cosine schedule:
% \[
\begin{equation}
\eta_{\text{t}}(t) = 0.6 - 0.4 \cos\left(\pi \frac{t}{T}\right),
% \]
\end{equation}

% \begin{equation}
% \eta_t = c + (1 - c) \cdot \cos^2\left(\frac{\pi}{2}\frac{T-t}{T}\right)
% \end{equation}
where $T$ is the total number of diffusion steps, and $t$ is the current timestep. The denoising update then becomes:
\begin{equation}
z_{t-1} = \sqrt{\alpha_{t-1}} \left( \frac{z_t - \sqrt{1 - \alpha_t} \epsilon_\theta(z_t, t)}{\sqrt{\alpha_t}} \right) 
+ \sqrt{1 - \alpha_{t-1} - \sigma_t^2} \cdot \epsilon_\theta(z_t, t) + \sigma_t \cdot \eta_t \cdot \epsilon,
\end{equation}
where $\sigma_t$ is diffusion step-size parameter and $\epsilon \sim \mathcal{N}(0, I)$. By progressively increasing $\eta_t$
  from 0.2 to 1.0, the schedule fosters stochastic exploration during translation while maintaining structural coherence. The integration of 
$\eta$-cosine noise schedule lowered FID scores (Table \ref{tab:image_translation_ablation} (right), $\eta$ cosine start 20\%), effectively striking a balance between morphological preservation and molecular fidelity.
\section{Experiments}
\begin{table}[t]
\caption{Evaluations on all translation tasks in the datasets, showing our superlative performance in almost all metrics. For PHV, $T=0.01$ and KIDs are multiplied by 1K.}
\label{tab:image_translationcomparison}
\centering
\renewcommand{\arraystretch}{1.2}
\fontsize{7.5pt}{9.5pt}\selectfont
\begin{tabular}{|c|c|c|c|c|c|c|c|c|c|c|c|}
\hline
Dataset & Method & SSIM$\uparrow$ & $PHV_{L1}$ $\downarrow$ & $PHV_{L2}$$\downarrow$ & $PHV_{L3}$$\downarrow$ & $PHV_{L4}$$\downarrow$ & MRA$\uparrow$ & FID$\downarrow$ & KID$\downarrow$ \\
\hline
\multirow{8}{*}{\begin{tabular}[c]{@{}c@{}}\(\mathrm{MIST}_{ER}\)\end{tabular}} & CycleGAN & 0.1982 & 0.5175 & 0.5092 & 0.3710 & 0.8672 & 0.235 & 125.7 & 95.1 \\
 % & CUT+$L{GP}$ & 0.2217 & 0.4531 & 0.4079 & 0.2725 & 0.8194 & 0.4882 & 43.7 & 8.7 \\
 & Pix2Pix & 0.1500 & 0.5818 & 0.5282 & 0.3700 & 0.8620 & - & 128.1 & 79 \\
 & PyramidP2P & 0.2172 & 0.4767 & 0.4538 & 0.3757 & 0.8567 & - & 107.4 & 84.2 \\
 & ASP & 0.2061 & 0.4336 & 0.4007 & 0.2649 & 0.8205 & 0.248 & 41.4 & 5.8 \\
 & MDCL & 0.2005 & 0.4533 & 0.3969 & 0.2646 & 0.8238 & 0.322 & 34.9 & \textbf{3.6} \\
 & InstructPix2Pix & 0.1536 & 0.5122 & 0.4961 & 0.3601 & 0.8679 & 0.210 & 61.3 & 38.9 \\
 & ControlNet & 0.2112 & 0.4992 & 0.4479 & 0.3370 & 0.8522 & 0.130 & 38.0 & 9.1 \\
 % & Confusion-GAN & - & - & - & - & - & - & - & - \\
 & \textbf{HistDiST} & \textbf{0.2278} & \textbf{0.3645} & \textbf{0.3105} & \textbf{0.2465} & \textbf{0.8226} & \textbf{0.552} & \textbf{31.3} & \textbf{3.6} \\
\hline
\multirow{5}{*}{\begin{tabular}[c]{@{}c@{}}\(\mathrm{MIST}_{HER2}\)\end{tabular}}
% & CycleGAN & 0.1914 & 0.5633 & 0.6346 & 0.4695 & 0.8871 & 0.6386 & 240.3 & 311.1 \\
 % & CUT+$L{GP}$ & 0.181 & 0.5321 & 0.4826 & 0.306 & 0.8323 & 0.5383 & 66.8 & 19 \\
 % & Pix2Pix & 0.1559 & 0.5516 & 0.507 & 0.3253 & 0.8511 & 0.5588 & 137.3 & 82.9 \\
 % & PyramidP2P & 0.2078 & 0.4787 & 0.4524 & 0.3313 & 0.8423 & 0.5262 & 104 & 61.8 \\
 & ASP & 0.2004 & 0.4534 & 0.4150 & 0.2665 & 0.8174 & 0.193 & 51.4 & 12.4 \\
 & MDCL & 0.1810 & 0.4371 & 0.3944 & 0.2518 & \textbf{0.8171} & 0.233 & 44.4 & 7.5 \\
 & InstructPix2Pix & 0.1353 & 0.5272 & 0.5045 & 0.3670 & 0.8740 & 0.125 & 65.6 & 33.7 \\
 & ControlNet & 0.1813 & 0.4437 & 0.4042 & 0.3070 & 0.8415 & 0.188 & \textbf{36.9} & 6.9 \\
 & \textbf{HistDiST} & \textbf{0.2059} & \textbf{0.3830} & \textbf{0.3298} & \textbf{0.2512} & 0.8245 & \textbf{0.466} & \textbf{36.9} & \textbf{3.8} \\
\hline
\multirow{5}{*}{\begin{tabular}[c]{@{}c@{}}\(\mathrm{MIST}_{Ki67}\)\end{tabular}} & ASP & 0.2410 & 0.4472 & 0.4001 & 0.2701 & 0.8128 & 0.148 & 51.0 & 19.1 \\
 % & CycleGAN & 0.3875 & 0.8274 & 0.8275 & 0.6081 & 0.9038 & 0.7917 & 343.9 & 317.9 \\
 % & CUT+$L{GP}$ & 0.1909 & 0.5426 & 0.4739 & 0.316 & 0.8415 & 0.5435 & 76.1 & 43.5 \\
 % & Pix2Pix & 0.1819 & 0.5468 & 0.4905 & 0.3415 & 0.8496 & 0.5571 & 147 & 142.4 \\
 % & PyramidP2P & 0.2286 & 0.4533 & 0.4222 & 0.336 & 0.8363 & 0.512 & 94.4 & 78 \\
 & MDCL & 0.2236 & 0.4005 & 0.3638 & 0.2465 & \textbf{0.8064} & 0.201 & 30.8 & 6.1 \\
 & InstructPix2Pix & 0.1756 & 0.5076 & 0.4894 & 0.3508 & 0.8578 & 0.145 & 63.0 & 32.6 \\
 & ControlNet & 0.2082 & 0.5205 & 0.4961 & 0.3712 & 0.8663 & 0.050 & 85.1 & 55.7 \\
 & \textbf{HistDiST} & \textbf{0.2463} & \textbf{0.3375} & \textbf{0.2893} & \textbf{0.2311} & 0.8174 & \textbf{0.484} & \textbf{28.05} & \textbf{4.4} \\
\hline
\multirow{5}{*}{\begin{tabular}[c]{@{}c@{}}\(\mathrm{MIST}_{PR}\)\end{tabular}} 
& ASP & 0.2159 & 0.4484 & 0.3898 & 0.2564 & 0.8080 & 0.237 & 44.8 & 10.2 \\
 % & CycleGAN & 0.2232 & 0.5334 & 0.5554 & 0.3867 & 0.8654 & 0.5852 & 96.1 & 96.6 \\
 % & CUT+$L{GP}$ & 0.2153 & 0.4656 & 0.4128 & 0.2724 & 0.8154 & 0.4916 & 54.6 & 20.1 \\
 % & Pix2Pix & 0.1617 & 0.6027 & 0.5569 & 0.4043 & 0.8601 & 0.606 & 183.8 & 148.1 \\
 % & PyramidP2P & 0.2403 & 0.5078 & 0.4682 & 0.3509 & 0.8446 & 0.5429 & 98.8 & 59.5 \\
 & MDCL & 0.2081 & 0.4320 & 0.3768 & \textbf{0.2499} & \textbf{0.8052} & 0.317 & 38.3 & 7.1 \\
 & InstructPix2Pix & 0.1339 & 0.5540 & 0.5387 & 0.3909 & 0.8786 & 0.126 & 82.7 & 60.4 \\
 & ControlNet & 0.2135 & 0.4291 & 0.3866 & 0.3039 & 0.8362 & 0.225 & 35.6 & 8.3 \\
 & \textbf{HistDiST} & \textbf{0.2381} & \textbf{0.4001} & \textbf{0.3355} & 0.2598 & 0.8265 & \textbf{0.434} & \textbf{33.7} & \textbf{4.3} \\
\hline
\multirow{5}{*}{\begin{tabular}[c]{@{}c@{}}\(\mathrm{BCI}_{HER2}\)\end{tabular}} 
% & CycleGAN & 0.4424 & 0.4264 & 0.3924 & 0.2610 & 0.7262 & - & 63.5 & 10.7 \\
 % & CUT+$L{GP}$ & 0.4802 & 0.4263 & 0.3784 & 0.2364 & 0.7328 & 0.4435 & 65 & 10.9 \\
 % & Pix2Pix & 0.4372 & 0.5121 & 0.4531 & 0.2953 & 0.7484 & - & 100 & 44.6 \\
 % & PyramidP2P & 0.5001 & 0.4531 & 0.3826 & 0.2618 & 0.7293 & - & 113.6 & 79.4 \\
 & ASP & \textbf{0.5032} & 0.4308 & 0.3670 & 0.2235 & 0.7210 & 0.058 & 65.1 & 9.9 \\
 & MDCL & 0.5029 & 0.4962 & 0.3861 & 0.2351 & 0.7344 & 0.084 & 51.2 & 13 \\
 & InstructPix2Pix & 0.3242 & 0.5575 & 0.5142 & 0.3804 & 0.8196 & 0 & 96.5 & 50.2 \\
 & ControlNet & 0.4581 & 0.5001 & 0.4233 & 0.2949 & 0.7577 & 0.07 & 43.5 & 10.2 \\
 & \textbf{HistDiST} & 0.4693 & \textbf{0.3136} & \textbf{0.2943} & \textbf{0.2165} & \textbf{0.7063} & \textbf{0.218} & \textbf{34.2} & \textbf{5.6} \\
\hline
\end{tabular}
\end{table}
\subsubsection{Datasets:}
We evaluate on two public H\&E-to-IHC datasets: (1) MIST ~\cite{li2023adaptive}: Paired H\&E patches ($1024 \times 1024$) with four IHC stains—HER2, Ki67, ER, and PR—using around 4k training pairs per stain and 1000 test pairs each. (2) BCI ~\cite{liu2022bci}: H\&E–HER2 pairs ($1024 \times 1024$) with 3896 training and 977 test samples.
% We evaluate our method using two public datasets used in H\&E-to-IHC stain transfer: \textbf{(1)  Multi-IHC Stain Translation (MIST) dataset} ~~\cite{li2023adaptive}: This dataset contains paired H\&E slide patches of size $1024\times1024$ with four IHC staining patches, including HER2 (4642 pairs), Ki67 (4361 pairs), ER (4153 pairs), and PR (4139 pairs) for training. We train our model for translating from H\&E to each of these four types individually and report evaluations over the available 1000 test pairs in the dataset for each combination. \textbf{(2)  Breast Cancer
% Immunohistochemical (BCI) challenge dataset} ~~\cite{liu2022bci}: This dataset contains pairs of H\&E and HER2 patches of size $1024\times1024$ provided in 3896 train and 977 test pairs.

\subsubsection{Implementation Details:}
We fine-tune Stable Diffusion v1.5, training for 500 epochs with a batch size of 16. AdamW optimization is used with a $2e-4$ learning rate, 1000 warmup steps, and a cosine decay schedule. Training alternates between H\&E generation and H\&E-to-IHC translation. $v$-prediction is set to $\gamma=5$, and noise schedule is rescaled to enforce zero terminal SNR. Lastly, during inference, we use 200 inversion steps and 200 denoising steps.
\subsubsection{Evaluation Metrics:}
We use both unpaired and paired metrics to assess image quality and molecular fidelity. Unpaired metrics include FID and KID for distributional similarity. Paired metrics include SSIM for image quality and PHV ~\cite{liu2021unpaired} for feature-level relevance. To capture pathology-specific semantics, we introduce Molecular Retrieval Accuracy (MRA), a novel metric leveraging GigaPath ~\cite{xu2024whole}, a pathology foundation model trained on H\&E and IHC data. MRA measures molecular fidelity by determining whether predicted IHC embedding best matches its ground-truth counterpart among all IHC embeddings in the test set using cosine similarity.
% We also report MRA@3 to account for biologically plausible variations.
% We assess our method using both unpaired and paired metrics to evaluate image quality and fidelity. As unpaired metrics, we compute Fréchet Inception Distance (FID) and Kernel Inception Distance (KID) to measure distributional similarity between generated and real images. For paired evaluation, we use common Structural Similarity Index (SSIM) and Peak Signal-to-Noise Ratio (PSNR) metrics. We also compute Perceptual Hash Value (PHV) at different layers of ResNet-101, following ~~\cite{liu2021unpaired}, to capture feature-level relevance. Additionally, to ensure the preservation of pathological semantics, we report accuracy (ACC) as the cosine similarity between ground-truth and generated patch features using GigaPath ~~\cite{xu2024whole}, a pathology foundation model trained on H\&E and multiple IHC stains.
% We also evaluate the molecular fidelity of predicted IHC images using the proposed Molecular Retrieval Accuracy (MRA) metric. MRA measures whether the predicted IHC embedding most closely matches the correct ground-truth embedding among all available candidates based on cosine similarity. We also report Top-3 MRA (MRA@3) to account for near-correct predictions reflecting biologically plausible variations.
\subsubsection{Results:}
We compare HistDiST with state-of-the-art stain translation models—CycleGAN ~\cite{zhu2017unpaired}, Pix2Pix ~\cite{isola2017image}, PyramidP2P ~\cite{liu2022bci}, ASP ~\cite{li2023adaptive}, and MDCL ~\cite{wang2024mix}—for H\&E-to-ER translation. As there are no specific diffusion-based approaches for H\&E-to-IHC stain transfer, we adapt two image translation methods widely used in the natural domain, InstructPix2Pix ~\cite{brooks2023instructpix2pix} and ControlNet ~\cite{zhang2023controlnet} and make comparisons with them. Specifically, we fine-tune InstructPix2Pix for our task, which is also used as the initialization for ControlNet. We train ControlNet with Canny edge map ~\cite{canny1986computational} of the H\&E input as the conditioning signal. For stain translation tasks on other staining types, we select the top two  GAN-based performers, along with InstructPix2Pix and ControlNet, for comparison. As shown in Table \ref{tab:image_translationcomparison}, HistDiST achieves consistent performance gains across all key metrics. It achieves the highest MRA scores, surpassing MDCL by 23\% on MIST\textsubscript{ER} and MIST\textsubscript{HER2}, demonstrating superior molecular fidelity. It also attains the lowest PHV scores, indicating better feature-space alignment, and the best FID/KID scores, confirming that the generated IHC images are more visually realistic and distributionally aligned with real ones.
% As presented in \ref{fig_qualitative} for H\&E-to-ER translation on the MIST dataset, HistDiST achieves superior performance compared to CycleGAN, ASP, and MDCL. Our method produces more realistic IHC correspondences, particularly in cases with inconsistent pairs. The various components of HistDiST contribute to enhanced patch quality, with $v$-prediction playing a key role in refining the color distribution to better match the target IHC domain. Additionally, applying inversion during inference helps preserve structural details by ensuring that the input noise is sampled from the latent H\&E distribution of our model. Finally, the scheduled injection of $\eta$-noise introduces controlled stochasticity, effectively addressing inconsistencies in paired data and further refining the color adaptation in the generated outputs.
\subsubsection{Ablation Studies and Qualitative Results:}
\begin{figure}[t]
    \centering
    \includegraphics[width=0.65\textwidth, angle=90]{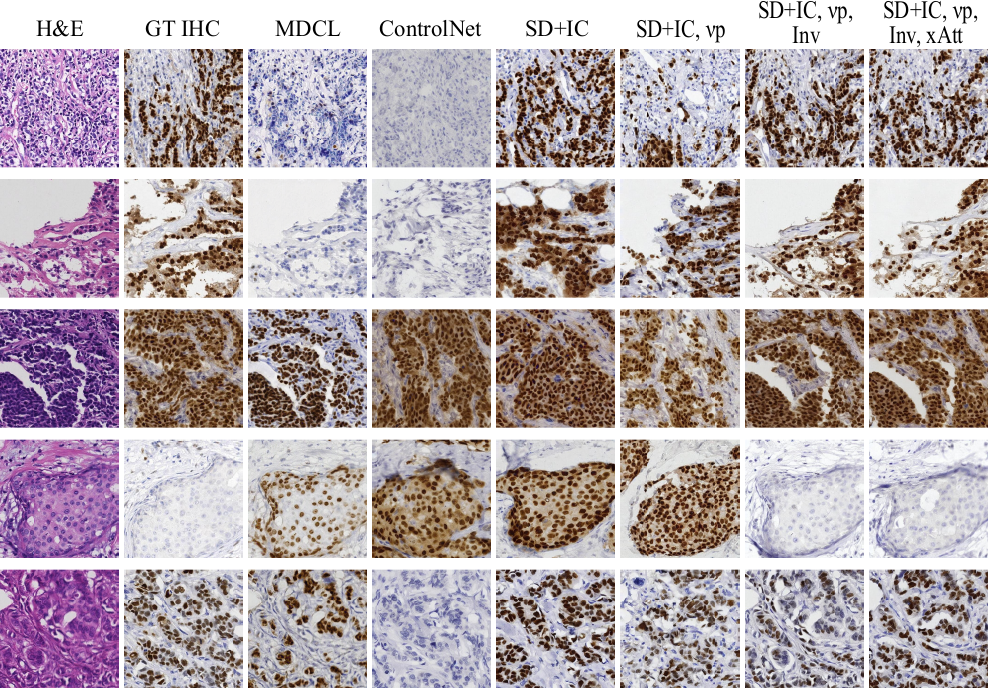}
    \caption{ER translation outputs across methods (see Table \ref{tab:image_translation_ablation} for abbreviations). The first two columns show the H\&E input and its corresponding ground-truth ER pair from the dataset, the next two columns (MDCL and ControlNet) show the previous top-performing GAN- and Diffusion-based approaches, and the last four columns are the incremental ablation of the components of our model, with our final model shown as the last column.}
    \label{fig:method_vis_comparison}
\end{figure}

\begin{table}[t]
\caption{Ablation study (on ER) during training (left) and inference (right). Abbreviations: "SD+IC": Stable Diffusion with input conditioning based on VAE-extracted H\&E features (InstructPix2Pix), "$v$-pred": $v$-prediction, "Inv": DDIM Inversion followed by DDIM sampling with $\eta$-cosine start $20\%$ noise, "xAtt": cross-attention with Phikon-extracted H\&E features. The inference strategies involve DDIM inversion for structure preservation, followed by DDIM sampling with different noise schedules (Fig. \ref{fig:inference} (right))}
\label{tab:image_translation_ablation}
\centering
\renewcommand{\arraystretch}{1.2}
\fontsize{8pt}{10pt}\selectfont
\begin{minipage}[t]{0.50\textwidth}
    \centering
    \begin{tabular}{|c|c|c|c|}
    \hline
    Training Strategy & SSIM$\uparrow$ & $PHV_{avg}\downarrow$ & FID$\downarrow$ \\
    \hline
    SD+IC & 0.154 & 0.559 & 61.3 \\
     % & CUT+$L{GP}$ & 0.4802 & 0.7328 & 65 \\
    SD+IC, $v$p & 0.198 & 0.501 & 37.5 \\
    SD+IC, $v$p, Inv& 0.228 & 0.511 & 41.3 \\
    SD+IC, $v$p, Inv, xAtt & 0.227 & 0.436 & 31.3 \\
    \hline
    \end{tabular}
    % \vspace{1ex}
    % \begin{center}
    % (a) Training stage ablation
    % \end{center}
\end{minipage}
\hfill
\begin{minipage}[t]{0.48\textwidth}
    \centering
    \begin{tabular}{|c|c|c|c|}
    \hline
    Inference Strategy & $PHV_{avg}\downarrow$ & MRA $\uparrow$ & FID$\downarrow$ \\
    \hline
     % & CUT+$L{GP}$ & 0.4802 & 0.7328 & 65 \\
    $\eta$ constant & 0.450 & 0.598 & 41.7 \\
    $\eta$ cosine  & 0.443 & 0.559 & 34.3 \\
    $\eta$ cosine start 20\% & 0.436 & 0.552 & 31.3 \\
    $\eta$ cosine start 30\% & 0.433 & 0.547 & 30.1 \\
    \hline
    \end{tabular}
    % \vspace{1ex}
    % \begin{center}
    % (b) Inference stage ablation
    % \end{center}
\end{minipage}
\end{table}
Table \ref{tab:image_translation_ablation} (Left) and Fig. \ref{fig:method_vis_comparison} present the quantitative and qualitative impact of different components in our framework. Fine-tuning Stable Diffusion with input conditioning alone on the paired dataset results in poor structural preservation and low molecular fidelity, leading to significant performance degradation across all metrics. Integrating $v$-prediction improves molecular expression patterns, enhancing FID scores and feature-space alignment. However, structural inconsistencies persist (as can also be seen in Fig. \ref{fig:method_vis_comparison}), indicating that direct adaptation remains insufficient for accurate stain translation.
\begin{figure}[t]
    \centering
    \includegraphics[width=0.2\textwidth, angle=90]{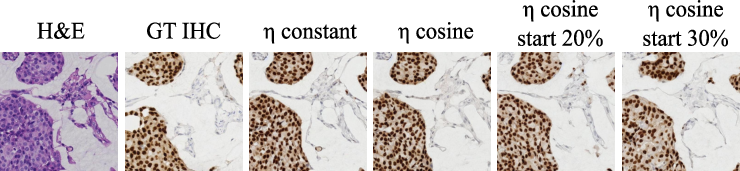}
    \caption{ER translation outputs across noise schedules (see Table \ref{tab:image_translation_ablation} for abbreviations).}
    \label{fig:schedules_vis_comparison}
\end{figure}
To address this, we conduct joint training on H\&E generation and H\&E-to-IHC translation, and at inference, apply DDIM inversion followed by DDIM sampling with $\eta$ cosine start 20\% noise schedule, which greatly enhances structural preservation. 
%However, this leads to high FID scores, suggesting a distributional shift in the generated IHC images. To counteract this, we introduce $\eta$-noise during DDIM sampling, which significantly reduces FID by enabling controlled stochasticity while maintaining structural consistency.
Finally, incorporating Phikon-extracted morphological embeddings into U-Net’s cross-attention layers further refines feature-space alignment and enhances structural integrity and molecular fidelity, yielding the best overall performance. This demonstrates that explicit pathology-aware conditioning is critical for accurate molecular reconstruction and high-fidelity stain translation. The qualitative comparison of our method against the best-performing GAN- and Diffusion-based approaches, MDCL and ControlNet, is also shown in Fig. \ref{fig:method_vis_comparison}. While ControlNet mostly provides improvements over the InstructPix2Pix baseline, it fails to preserve the structure accurately since acquiring Canny edges for low-contrast H\&E images is challenging, and also fails to obtain accurate IHC expressions. Furthermore, in most cases, MDCL often introduces artifacts or fails to capture fine molecular details. On the other hand, HistDiST better preserves molecular expression patterns and produces more precise IHC intensity distributions, aligning closely with GT IHC. For instance, in the fourth row of Fig. \ref{fig:method_vis_comparison}, MDCL and ControlNet incorrectly predict the molecular expression pattern (shown with a high expression level) while our method accurately predicts the patterns as low expression levels (last column).

Table \ref{tab:image_translation_ablation} (right) and Fig. \ref{fig:schedules_vis_comparison} examine the effect of different $\eta$-noise schedules at inference, with Fig. \ref{fig:inference} (right) visualizing the tested schedules. Setting $\eta=0$ yields optimal structural alignment between input H\&E and translated IHC but results in high FID scores, indicating a poor match to the real IHC distribution. Introducing a cosine noise schedule starting from zero improves FID scores while preserving structural integrity, suggesting that injecting controlled stochasticity at higher timesteps promotes distributional alignment with real IHC data. Increasing the starting noise level to 0.2 further reduces FID scores with minimal structural loss, while increasing it to 0.3 introduces noticeable structural deviations, despite additional improvements in FID. To balance structural preservation, molecular fidelity, and distributional realism, we adopt the cosine schedule starting at 0.2 as the optimal configuration.

% Our ablation studies provide experiments with various training-based and inference-based combinations of our components separately. \ref{table_ablation_train} illustrates the effectiveness of our training-phase components. We observe that the joint training of H\&E generation and H\&E-to-IHC translation using diffusion improves the results, as denoising happens in a more coherent latent space within the underlying UNet. Furthermore, this setup enables the employment of inversion noise later during inference, since the inversion sample needs to correspond to the H\&E latent space, which is not guaranteed by solely training the IHC translation task. Furthermore, as depicted by unpaired metrics in the table, $v$-prediction/offset noise improves the fidelity of the generated images.

% \input{sections/5_conclusion}
\section{Conclusion and Limitations}
We present HistDiST, an LDM-based framework for H\&E-to-IHC stain translation, integrating pathology-aware conditioning, DDIM inversion, and zero terminal SNR enforcement to enhance structural fidelity and molecular accuracy. Notably, the inversion process helps to preserve the structure; still, we employ controlled stochasticity during denoising as fully-deterministic sampling negatively affects the fidelity of generated images. We also introduce Molecular Retrieval Accuracy (MRA), a novel metric for pathology-aware evaluation based on the alignment of pathological features. Experiments on MIST and BCI datasets demonstrate that HistDiST achieves superior visual fidelity, molecular accuracy and structural preservation compared to the existing approaches. Nevertheless, we emphasize that this method is not clinically applicable at this stage; additional validation and analysis (e.g., expert review or downstream task performance) would be required before considering diagnostic use. Future studies can further explore the potential of stain transfer methods by benchmarking their utility in downstream tasks such as patch-level classification or segmentation.

\section{Acknowledgments}
The work described in this paper was conducted in the framework of the Graduate School 2543/1 “Intraoperative Multi-Sensory Tissue Differentiation in Oncology" (project ID 40947457) funded by the German Research Foundation (DFG - Deutsche Forschungsgemeinschaft). This work has been supported by the Deutsche Forschungsgemeinschaft (DFG) – EXC number 2064/1 – Project number 390727645. The authors thank the International Max Planck Research School for Intelligent Systems (IMPRS-IS) for supporting Valay Bundele and Mehran Hosseinzadeh.
% Note: if you want to use up all of the allowed space for the paper,
%       the bibliography will start on top of page 13. Furthermore,
%       from page 13 onwards, there will be *only* bibliography, no more
%       figures/tables.
\bibliographystyle{splncs04}
\bibliography{055-main}

\end{document}